\documentstyle[twocolumn,aps]{revtex}
\newcommand{\be}{\begin{equation}}
\newcommand{\ee}{\end{equation}}
\begin{document}
\title{ The Light--Cone nonperturbative dynamics \\
            of meson wave functions }
\author{V.L.~Morgunov, V.I.~Shevchenko, Yu.A.~Simonov }
\address{Institute for Theoretical and Experimental Physics \\
  RU-117259,  Moscow,  Russia} 
\date{\today}
\maketitle
\begin{abstract}
The light--cone Hamiltonian, incorporating the nonperturbative
dynamics of the $q\bar q$ system connected by the string is solved
numerically. The spectrum is shown to coincide with that of the
center--of--mass Hamiltonian within the accuracy of computation,
displaying the expected degeneracies of the string states, but
an overall shift due to different treatment of Z--graphs is obtained.
The nonperturbative wave functions are calculated directly from
the light--cone Hamiltonian for the first time, thus allowing the explicit
estimation of nonperturbative effects in the parton model language.
In this way one obtains the string contribution to the parton distributions
and in the formfactors and structure functions.
\end{abstract}

\section{Introduction}

The light--cone description of the hadron wave functions
is widely used, since it allows to get a direct connection to the parton
model and its QCD improvements \cite{feynm}. The latter however are mostly
perturbative and nonperturbative contributions are introduced
via OPE and QCD sum--rules. In this way the concept of the string -- the
main nonperturbative QCD phenomenon -- is totally lost. On physical grounds
it seems that the string is the essential ingredient of the dynamics
for large (${r\ge0.5\>fm}$) distances, and it would be interesting to understand
its contribution to the light--cone wave functions, formfactors, structure
functions etc.

In particular, what is the QCD string in the parton language? Should one
associate it with the gluon contribution as an assembly of gluons
compressed inside the string -- or with the constituent quark mass?

There two contrasting points of view have been proposed already decades ago
\cite{kutwei,altar1}.
In \cite{kutwei} the sea quarks and gluons enter as separate entities
and one could associate the gluon distribution with the string
in the same way as photons with the Coulomb field of the charge
in the Williams-- Weizs\"acker method. In contrast to that
in \cite{altar1} the quarks have been considered as constituents with
structure, and string does not appear separately.
Recently a quantitative analysis was performed \cite{altar2}
of quark distribution in the pion starting from that in the nucleon
and assuming the same internal structure of quarks in nucleon and in pion.
It is still an open question how the structure of the constituent quarks
is formed, and to which extent it can be explained by the adjacent piece
of the string.
This problem can be elucidated partly in the present
approach, since our light--cone Hamiltonian contains the string on the light
cone explicitly.
It allows to separate the contribution of the string to the
parton distribution, in particular to the momentum sum--rule.
There is another source of structure in the constituent quark -- chiral
symmetry breaking which creates the chiral mass of quark.
This problem will not be discussed below, see e.g. \cite{sim1}.

There are many other questions which can be asked from a decent light--cone
Hamiltonian, incorporating $q\bar q$ system with the proper light--cone
nonperturbative dynamics.
An important advantage of the light--cone wave functions is that they allow
to calculate formfactors and structure functions directly, without additional
boosts, which typically are the dynamical ones, i.e. require the use of the
Hamiltonian.

Another point to clarify is the comparison of the proper light--cone dynamics
incorporated in the wave--function of the light--cone Hamiltonian
\cite{dub1} with the popular anzatz where one is simply using the c.m.
wave function expressing it through the light--cone variables \cite{chung}.

But the first question to answer is the comparison of the c.m. and light-cone
spectra. In the c.m. system the Hamiltonian of the spinless $q\bar q$
system, connected by the minimal (i.e. without vibrations) string was
obtained in \cite{dub2}.
The backward--in--time motion or so called Z--graphs are suppressed
there because Z--graphs contains backward moving string
and the corresponding increase in the action damps the amplitude.
On the light--cone Z--graphs are treated differently and do not appear in
lowest orders of perturbation theory
and comparing spectrum with that of c.m. one can visualize
the difference due to the different contribution of Z--graphs.

Another typical feature of the light--cone Hamiltonian should be mentioned.
As we shall see in Section~2 the form found in \cite{dub1} explicitly contains
the z--axis component of the angular momentum, $L_z$ which is not
Lorenz--invariant and hence the light--cone spectrum should demonstrate
degeneracy for different values of $L_z$ with the same $L^2$.
This degeneracy will be explicitly demonstrated below.
Therefore all states of the system including angular excitations may
be calculated using the Hamiltonian with $L_z = 0$.
It is known (see \cite{dub2} and also \cite{cours})
that the problem under consideration possesses the dynamical symmetry
which implies that quasiclassical frequencies of the radial and orbital
motions are approximately proportional to each other with the coefficient 2.
Thus the light--cone Hamiltonian should also manifest this dynamical symmetry.

The plan of the paper is as follows. In Section~2 we start with
the light--cone Hamiltonian in 3+1--dimensions from \cite{dub1} and
prepare it
for numerical computations. As in \cite{dub1}, we simplify the matter
considerably, neglecting spins, perturbative gluon exchanges and additional
quark pairs. The first two points are unessential at large distances,
which are of the main interest for us here but seriously contribute to
the lowest mass mesons. We plan to include these effects in next publications.
Sea quarks are expected to contribute some 10\% effect and disappear
in the large $N_c$ limit. As it is, we are confined to the valence
quark sector of the Fock column wave function.

In Section~3 general properties of solutions and the physical
meaning of parameters ${\mu}_i , x$ and $\rho$ are discussed.
We also demonstrate the limiting procedure reducing the 3+1 Hamiltonian
to the 1+1 problem,
following the method of \cite{dub1} and compare properties of 3+1 and 1+1
solutions. In Section~4 the formfactors and structure functions are
defined through the solutions of the light--cone Hamiltonian and physical
limiting cases are discussed.
In Section~5 the numerical procedure used for the solution of the eigenvalue
equation is explained. In Section~6 we present numerical results and discuss
them from different points of view, e.g. comparing center--of--mass spectrum
and wave functions with those of the light--cone.
Special attention is paid to the procedure used in literature \cite{chung}
where the c.m. wave function is kinematical expressed in terms of the
light--cone variables.
A short conclusion and perspectives
are given in Section~7. Two appendices, A and B serve to illustrate
the derivation of the Hamiltonian (\ref{ham}) and the final form of equation
(\ref{seq}) to be solved numerically.

\section{The Green's function and the Hamiltonian}

We consider the relativistic quark--antiquark pair with the masses
$m_1$ and $m_2$ connected by the straight--line Nambu--Goto string with
the string tension $\sigma$ in 3+1 dimensional space--time, see \cite{dub1}
and Appendix A of the present paper for details.
We start from the expression for the classical light--cone Hamiltonian function
in 3+1 dimensions
\begin{eqnarray}
 H=\frac{1}{2} \left\{ \frac{m^2_1}{\mu_1}+\frac{m^2_2}{\mu_2}+
 \frac{L_z^2}{a\>r^2_{\bot}}+
 \frac{(p_{\bot}r_{\bot}+\gamma r_-)^2}{\tilde{\mu} r^2_{\bot}}\right.
\nonumber \\
+\left.\int \frac{\sigma^2}{\nu} d\beta r^2_{\bot}+ \frac{\nu_0 P_+}{\mu_1+\mu_2}
 \frac{r^2_-}{r^2_{\bot}}  \right\}
\label{ham}
\end{eqnarray}
In Appendix A we briefly derive this expression, using \cite{dub1} and
omitting unessential details.
Here $\mu_1, \mu_2 $ are einbein fields, playing the role of $P_{+}$
momenta of the particles, $\tilde{\mu} =
\mu_1 \mu_2 / (\mu_1 + \mu_2)$
,  $\nu_0 = \int_{0}^{1} d\beta \nu(\beta)$ and $\nu(\beta)$ is
the einbein field with the physical meaning of the $P_{+}$--momentum,
carried by the string.
As it has been mentioned above, the Hamiltonian explicitly depends
on $L_z^2 = {\vec p}~^{2}_{\bot}{\vec r}~^{2}_{\bot} -
({\vec p}_{\bot}{\vec r}_{\bot})^2
$, the corresponding mass parameter in the denominator in (\ref{ham})
is equal to
\be
a=\mu_1(1-x)^2 + \mu_2 x^2 + \int\limits_0^1 d\beta \nu(\beta)
(\beta - x)^2
\label{adef}
\ee
The variable $x$ in the above expression is defined as (see Appendix A
for details):
\be
x=\frac{\mu_1 + <\beta>\nu_0}{P_+}
\label{x}
\ee
where
\be
 P_+ = \mu_1 + \mu_2 + \nu_0
\label{P}
\ee
is the light--cone total momentum of the system.
Also
\be
\gamma = \nu_0\left(<\beta> -
\frac{\mu_1}{\mu_1 + \mu_2}\right)
\label{gamm}
\ee
where $<\beta> = \int_0^1 d \beta \beta \nu(\beta) / \int_0^1 d \beta
\nu(\beta)$.

The corresponding Green's function is defined as an integral over the dynamical
fields as well as over the einbein fields:
\be
G(x\bar{x};y\bar{y})= \int
D{\mu}_1D\mu_2D\nu DR_{\mu}Dr_{\mu}e^{-A}
\label{green}
\ee
where $A$ is the euclidean
action. Separating out the center of mass motion one defines the Hamiltonian
from the
 corresponding Minkowskian action $A^M$:
\be
A^M =  \int dz_+ L^M~, ~~H=p_{\bot}\dot{r}_{\bot}-L^M
\label{mink}
\ee
with $p_{\bot} = \partial L^M / \partial \dot{r}_{\bot}$.
This is the light--cone Hamiltonian of our problem (\ref{ham}).

As the next step one should quantize the classical Hamiltonian function.
Before doing it,
one should choose the appropriate set of dynamical variables.
Three einbein fields ${\mu}_1 , {\mu}_2 , \nu $ introduced above play
different dynamical roles. In the
nonrelativistic case $m_1, m_2 \gg \sqrt{\sigma}$ (and therefore for the free
particles) the dependence of Hamiltonian (and wave functions) on $\nu$ can be
correctly found by minimization procedure with $\nu$ taken as a classical
variable.  This is in its turn a consequence of the fact, that string in our
approach is the minimal one -- it has no internal degrees of freedom and may
only stretch or rotate as a whole.

On another hand, $\mu_1$ and $\mu_2$ on the light cone play the role
of legitimate quantum dynamical degrees of freedom and can be expressed
through $x$ and $P_+$ as in (\ref{P}) and (\ref{x}).

There are two canonically conjugated pairs
$\{{\vec p}_{\bot}, {\vec r}_{\bot}\}$ and $\{x, (P_+r_-)\}$ (see Appendix A).
We introduce also a new
dimensionless variable $\tilde y$ instead
of $\nu$ : $\tilde y (\beta) = {\nu}(\beta)/{P_+}$.
It satisfies the obvious condition:
$0<\tilde y<1$. This variable depends on ${\vec r}_{\bot}$ as well as on
$(P_+r_-)$.
Rigorously speaking one should extract this dependence by the minimization
of the Hamiltonian with respect to $\tilde y$ as it has been explained above.
Instead an easier (however approximate) way is chosen. On physical
grounds it can be shown that for small
$r^2_{\bot}$ one has ${\tilde y}_0 = \int_0^1 \tilde y d\beta \sim
const\cdot r^2_{\bot}$ so that linear string energy density
${\tilde y}_0/r^2_{\bot}$ stays finite if $r_{\bot}^{2} \to 0$, 
and using this property one can
reproduce the correct 1+1 limit, namely 't Hooft equation \cite{thof},
from the 3+1 light--cone Hamiltonian (\ref{ham}) (see \cite{dub1} and
the next section).
On the other hand, if distances are increasing, ${\tilde y}_0$ tends to some
limiting value which is determined by the virial theorem arguments.
So one can parameterize ${\tilde y}$ introducing several parameters
and replace the minimization in the functional sense by the ordinary
minimization with respect to these parameters.
We have chosen the simplest 2--parametric form
for ${\tilde y}$:
\be
\tilde y = \frac{yt}{1+\alpha t}
\label{y}
\ee
where $t={r^2_{\bot}}$ and $y$ and $\alpha$ are free parameters.
The requirement $0 < \tilde y < 1$ leads to the restriction
$0 < y < \alpha $. Let us stress again that this parameterization is
the matter of convenience and all physical results are determined
from the requirement that every energy level should have its own minimum.

It is easy to see from (\ref{x}), that the $x$--variable  is
the part of the total momentum, carried by the first quark itself and a
part of the string, "belonging" to this quark.
Rewriting (\ref{x}) in the form:
\be
\frac{\mu_1}{P_+} = x - \tilde y <\beta> \;\;;\;\;\frac{\mu_2}{P_+} = (1-x) -
\tilde y (1-<\beta>)
\label{muP}
\ee
one can conclude, that for the given $\tilde y$ the variable $x$ may vary in
the following limits:
\be
\tilde y <\beta>\; \le\; x\; \le\; 1 - \tilde y \>(1 - <\beta>)
\label{yt}
\ee
It is more convenient to make rescaling to a new variable $\rho $ which
varies from zero to unity:
\be
x = \tilde y <\beta> + (1-\tilde y) \rho\;\; ; \;\; 0\le\rho\le 1
\label{r}
\label{xt}
\ee
The quantity we are really interested in is the mass operator squared:
\be
{\hat M}^2 = 2\>{\hat H}\>P_+
\label{msq}
\ee
and for ${\hat M}^2$ we have the Schr\"odinger
equation of the following form:
\be
A_1\>\psi '' + A_2 \ddot \psi + A_3 \dot \psi ' +
 A_4 \psi ' + A_5 \dot \psi + A_6 \psi = M^2 \psi
\label{seq}
\ee
here $\psi = \psi(\rho, {\vec r_{\bot}})$ and two independent
variables are $\rho\in [0,1]$ and
$t=r^2_{\bot}\in
[0, \infty)$ ; $\psi' \equiv \partial \psi / \partial t ;
\dot \psi \equiv \partial \psi / \partial \rho $ ;
the coefficients $A_i$ depend on $\rho$ and $t$.
The derivation of (\ref{seq}) from the {Hamiltonian (\ref{ham})}
as well as the explicit form of the functions $A_i$
may be found in Appendix B.
The following boundary conditions are to be imposed:
\be
\psi(0, t) = \psi(1, t) = 0 \;\; ,\;\;\; \psi(\rho, t\to\infty) = 0
\label{psi0}
\ee
with the normalization of the solution:
\be
\int\limits_0^1d\rho \int d^2 {r_{\bot}} |\psi(\rho, {\vec r_{\bot}})|^2 = 1
\label{n}
\ee
We discuss properties of solutions of (\ref{seq}) in the next section.

\section{General properties of the solutions.}

In this section we shall discuss the physical meaning of our variables
$x, \mu, \rho$ and $\tilde y$ and some properties of the solution
$\psi(\rho, t)$. Rewriting (\ref{P}) as
\be
1=\frac{\mu_1}{P_+} + \frac{\mu_2}{P_+} + \tilde y
\label{one}
\ee
one can conclude that $\mu_i / P_+$ and $\tilde y$ have the meaning of
the parts of the total momenta $P_+$, associated with the $i$--th quark
and string respectively. This identification is also supported by the
form of the Hamiltonian (\ref{ham}), where masses enter as
$m_i^2 / 2 \mu_i$
so that in the infinite--momentum frame it becomes
$m_i^2 / 2p^i_{+}$.
The similar situation occurs in the c.m. system where $\mu_i$ and $\nu$
have the meaning of the energy, associated with the quarks and the string.
One notes the essential difference in the way these einbein fields
are treated in the c.m. Hamiltonian and on the light--cone. In the former
case $\mu_i$ enter on the same grounds as $\nu$ (or $\tilde y$) and are to be
found by minimization of the Hamiltonian.  In contrast to that on the
light--cone $\mu_1, \mu_2$ are expressed through $\tilde y$ and $x$ as
in (\ref{x}) and $x$ is the canonical momentum, conjugated to the coordinate
$P_+ r_-$ ( see Appendix A), namely:
\be
[x, P_+r_-] = -i
\label{conj}
\ee
Hence $\mu_i$ are promoted to be quantum operators with average values
to be found from the solution of the "Schr\"odinger equation" (\ref{seq})
Without the string, when $\tilde y = 0$ one has
$\mu_1 = xP_+ \;\; ; \;\; \mu_2 = (1-x) P_+$ and the correspondence
with the naive parton model is trivial.
However when $\tilde y \ne 0$ there are no {\it a priori } reasons to
associate our $x$ with the Feynman $x$ parameter, but instead one could
call $\mu_i / P_{+}$ the Feynman $x$ parameter and associate $\tilde y$
with the Feynman parameter for the string.

However the strong interaction between the string and quarks taken into
account in the light--cone Hamiltonian (\ref{ham}) invalidates the literal
interpretation of our results in the free parton language -- one can only
interpret the averaged values of $\mu_1, \mu_2$ together with the minimized
value for $\tilde y$.

Then one can consider the momentum sum--rules for the quark structure functions
$f_2^F(x_F)$ where the quark Feynman parameter $x_F$ normalized to the
interval [0,1] should be associated with the parameter $\rho$ from (\ref{xt})
which is connected with $\mu_i$ as follows:
\be
\frac{\mu_1}{P_+} = (1-\tilde y)\rho \;\; ;\;\;
 \frac{\mu_2}{P_+} = (1-\tilde y)(1-\rho)
\label{mu1}
\ee

Therefore as we shall discuss later the average momentum carried by the
quarks is equal to $(1-\tilde y)$ while the rest, $\tilde y$ is carried
by gluons assembled into the string -- nonperturbative gluons.
This interpretation of the missing momentum (around one half in
experiment \cite{feynm} ) as due to the string was suggested already
in \cite{kutwei}.
Now we are in the position to calculate this quantity
and compare to the experiment which we shall do in Section~6.
As we shall see, the main missing ingredient in our Hamiltonian  and wave
functions is the sea quarks and "sea gluons" -- i.e. gluons which excite
the string and make it vibrating instead of keeping the straight--line form
we have considered. From this point of view our wave function, the solution of
(\ref{seq}) is the first upper entry in the Fock column, containing states
with 1,2,... gluons, "sitting on the string" and exciting it to the first,
second, etc. vibrating string state. It is the whole Fock column which
is responsible for the Regge behaviour of cross--sections and Regge asymptotic
of structure functions near $\rho \to 0$. In this paper we concentrate
on the properties of the sector with two valence quarks and minimal
(i.e. nonvibrating) string, leaving the general discussion of the QCD
reggeons and the pomeron to a future publication.

Let us now turn to the properties of the wave functions and, in particular,
their behaviour at the boundaries. Inserting in (\ref{seq}) $\psi(\rho)
\sim \rho^{\alpha}$ and $(1-\rho)^{\alpha}$ near the points $\rho = 0$
and $1$ respectively one obtains the only solution $\alpha = 1$ at all $t$
hence the boundary conditions (\ref{psi0}) are satisfied.
Similarly,  representing $\psi(\rho, t=r_{\bot}^2)$ at large $t$ in the form
\be
\psi(\rho, t) \sim A(\rho) exp(-\gamma(\rho) t^{\xi})
\label{psit}
\ee
yields $\xi =1$ and some complicated equation for $\gamma(\rho)$.

Finally, we discuss in this section the limit of the (1+1) 't Hooft equation
\cite{thof}. Following the arguments given in \cite{dub1} we look for the
limit $t\to 0$ with $\tilde y = yt/(1+\alpha t) \to 0$ and $y$ fixed
and to be found from the minimization. As a result, one obtains:
\be
H_1 = \frac12\left\{ \frac{m_1^2}{\mu_1} + \frac{m_2^2}{\mu_2} +
\int\frac{{\sigma}^2}{y(\beta)} d\beta +
\int y(\beta) d\beta \frac{P_+ r_-^2}{\mu_1 + \mu_2} \right\}
\label{H1}
\ee

Note that the string part in the sum (\ref{one}) disappears because of
$\tilde y = 0$ but $y$ is nonzero and ensures the string--like
potential dynamics.
Taking the extremum of (\ref{H1}) with respect to $y(\beta)$ and
introducing $x$ instead of
$\mu_1, \mu_2$ where $\mu_1 = P_+x, \mu_2 = P_+(1-x)$
one obtains:
\be
M^2 =2P_+H_1 = \frac{m_1^2}{x} + \frac{m_2^2}{1-x} + 2\sigma |P_+r_-|
\label{M}
\ee
From (\ref{M}) one easily deduces the 't Hooft equation \cite{thof}:
\be
\left( \frac{m_1^2}{x} + \frac{m_2^2}{1-x}\right)\psi(x)
-\frac{\sigma}{2\pi}\int\frac{dy \psi(y)}{(x - y)^2} = M^2 \psi(x)
\label{thth}
\ee
Thus the 't Hooft solutions \cite{thof} which approximately are:
\be
\psi_n(x) = C_n \sin \pi n x
\label{psin}
\ee
can be thought of as limiting case $t\to 0$ of our solutions but one should
have in mind that the extremum value of $y$ leading to the 't Hooft
equation (\ref{thth}) in general does not coincide with the extremum value
in our equation (\ref{seq}) obtained for all nonzero $t$. Still the
behaviour at the boundary points $x=0,1$ ($\rho = 0,1$ in our case)
and the general pattern of our solution $\psi(\rho, t=0)$ is closely
resembling those of 't Hooft solutions.

\section {Formfactors and structure functions}

The formfactor of the bound system of two spinless quarks can
be expressed through the wave--function in the momentum representation,
depending on $x$ and on the relative momentum $p_{\bot}$ as follows
(for equal mass quarks) \cite{drell}:
\be
F(q^2) = \int \psi(x, {\vec k}_{\bot}) \psi^*(x, {\vec k}_{\bot} +
(1 - x)\vec{q}_{\bot}) dx d^2 k_{\bot}
\label{Ff}
\ee
Our wave function is defined in the mixed space $\rho$, $r_{\bot}^2 = t$
hence one obtains:
\begin{eqnarray}
F(q^2) = \pi \int\limits_{0}^{\infty} \int\limits_{0}^{1}
|\psi(\rho, t)|^2 \nonumber \\
J_0 \left\{ q\sqrt{t} \;
\left[1 - \rho + \tilde y (\rho - \frac{1}{2}) \right] \right\} d\rho dt
\label{Ff1}
\end{eqnarray}
where we have exploited relation (\ref{xt}) with $<\beta> = 1/2$.
For $q^2 = 0 $ the usual normalization follows $F(0) = 1$ since our wave
function is normalized as
\be
\pi \int\limits_{0}^{\infty} \int\limits_{0}^{1}
|\psi(\rho, t)|^2 d\rho dt = 1
\label{Fnorm}
\ee

Note that the effect of the electromagnetic current, which interacts
only with the charged quark (and antiquark) and does not interact
directly with the string, enters (\ref{Ff1}) through the factor
$(1 - x) \vec{q} = [1 - \rho + \tilde y (\rho - 1/2) ]\vec{q}$,
and not through $\mu$ or $\rho$. This is
the result of the fact, that the parameter $x$ governs the
distribution of the momentum between the center of mass and
relative coordinates $\stackrel{.}{R}_{\bot}$ and $\stackrel{.}{r}_{\bot}$.

 Let us notice, that because we neglect perturbative gluon exchanges,
our formfactor does not have the familiar quark asymptotic
$ F(q^2) \sim const / q^{2}$ for $ q^2 \to \infty$, but rather
corresponds to the nonperturbative part which is believed to dominate
the low--$q$ region (actually the experimentally accessible region
of today) \cite{radush}.

From (\ref{Ff1}) and the behaviour of $\psi(\rho, t)$ at large $t$ and
$\rho \to 1 $, equation (\ref{psit}), one easily obtains that
\be
F(q^2) \sim  \frac{const}{|q|^3} \; \; \; ,\; |q| \to \infty
\label{Fto}
\ee

We turn now to the structure function (or quark distribution function).
As well as in case of the formfactor, since as it is produced by the
external current it is natural to define it a function of $x$, but to
compare it with standard results normalized to the interval [0,1], we
choose to define it as a function of $\rho$.

Thus, the parton distribution inside our meson wave function can be written as:
\be
q(\rho) = \pi \int\limits_{0}^{\infty}  |\psi(\rho, t)|^2 dt
\label{qr}
\ee
and the normalization of $q(\rho)$ due to (\ref{Fnorm}) is
\be
\int\limits_{0}^{1} q(\rho) d\rho = 1 \;.
\label{qnorm}
\ee
The momentum sum--rule, using (\ref{one}) and (\ref{mu1}) is
(for equal masses of quark and antiquark)
\be
2 \int\limits_{0}^{1} q(\rho) \rho (1 - <\tilde{y}>) d\rho +
\int\limits_{0}^{1} q(\rho) <\tilde{y}> d\rho = 1
\label{qnorm1}
\ee
where $<\tilde{y}>$ is obtained by integration with $|\psi(\rho, t)|^2$,
\be
<\tilde{y}> = \pi \int\limits_{0}^{\infty} \int\limits_{0}^{1}
\tilde{y} \; |\psi(\rho, t)|^2 d\rho dt
\label{ytil}
\ee

Comparing with the standard energy--momentum sum--rule \cite{feynm}
containing the gluon distribution $g(x)$, one may associate it with
$\tilde y$ as follows $\int^{1}_{0} g(x) x dx \Rightarrow <\tilde y>$ .

From (\ref{qnorm1}, \ref{yt}) one can see that both valence quarks
together carry the part of momentum $P_{+}$ equal to $1-<\tilde{y}>$,
while the string carries on average $<\tilde{y}> P_{+}$ .

The essential difference of the minimal string from the gluon distribution
$g(x)$ is that the former is fixed in the quasiclassical einbein
formalism and has no dispersion -- unlike free gluons which can carry
any part of the total momentum $P_{+}$ .

Let us discuss finally the correspondence between $q(\rho)$ and $F(q^2)$
(the Drell-Yan-West duality \cite{west}).
If $F(q^2)$ behaves as $(q^2)^{-n}$ at large $q^2$, then $F_2(x)$ should
decrease as $(1-x)^p$ for $x \to 1$, where $2 n = p + 1$.
In our case from (\ref{Fto}) $n = 3/2$, hence $p=2$  and this agrees with
(\ref{qr}), since $\psi(\rho) \sim (1-\rho)$, and $q(\rho) \sim (1-\rho)^2$.

\section {Numerical solution}

The equation (\ref{seq}) is the linear differential equation of the elliptic type.
Boundary conditions are given in (\ref{psi0}) and are compatible with the
singularities of coefficients $A_{i}$. We have not succeeded in finding any
analytic function, describing the main features of the solution,
for example, one can
demonstrate that simple ansatz $ \psi(\rho, t) = \phi(\rho)\cdot
\xi(t) $ is not a solution of (\ref{seq}). Therefore we admitted
two numerical procedures: a direct solution of (\ref{seq})
using discretization and a variational ansatz with the expansion of
the solution into a complete set of functions of two variables.
We have found that the former procedure turned out to be unstable and not
accurate. We discuss below only the numerical results obtained
in the framework of the second method.

The $\Psi$ function in (\ref{seq}) depends on parameters $m_1 , m_2$
(bare quark masses), $\sigma$ (string tension), 2 parameters $y$
and $\alpha$ defining the string parameter $\tilde y$ (see (\ref{y})).
The $\Psi$ also depends on quantum numbers: $L_z$, total momentum $L$
and radial quantum number $N_r$.
The eigenvalue of (\ref{seq}) is the bound state mass squared
$M^2(L_z, L, N_r , y, \alpha)$. We expand $\Psi$--function into the
complete set of functions ($\kappa = L_z, L, N_r$):
\be
\Psi_{\kappa}(\rho, t) = exp(i L_z \phi) \sum\limits_{m=1}^{\infty}
\sum\limits_{n=0}^{\infty} C^{\kappa}_{mnl} {\psi}_{mnl}
\label{yyy}
\ee
where
\be
\psi_{mnl}(\rho, t) =  \sin(\pi m\rho) \;
 t^{\frac{l}{2}}\;
L_n^{l}(\epsilon t) \exp(-{\epsilon t}/{2}).
\label{yy}
\ee
Here $\phi$ is the azimuthal angle of the corresponding angular motion in the
$xy$--plane and $L_n^{l}$ -- Laguerre polynomials.
The variational parameter
$\epsilon$ and the definition $l\equiv L_z$ were also introduced.
The basis functions are orthonormalized as:
\begin{eqnarray}
\int\limits_0^{\infty}\frac{dt}{2}\int\limits_{0}^{2\pi}d\phi
\int\limits_0^1 d\rho \psi_{m'n'l'}^{*}(\rho, t) \psi_{mnl}(\rho, t) =
\nonumber \\
= {\delta}_{m'm}\>{\delta}_{n'n}\>{\delta}_{l'l}\>\frac{\pi}{2
{\epsilon}^{l+1}}\frac{(n+l)!}{n!}
\label{psinorm}
\end{eqnarray}
and the total function is normalized as:
\be
\int dV |\Psi|^2 = 1 \; \; \; dV = \frac{1}{2} d \rho dt d \phi
\label{psinorm1}
\ee
Note, that each function ${\psi}_{mnl}$ satisfies boundary conditions
(\ref{psi0}). We have used typically seven sine functions $(m=1,..,7)$
and seven Laguerre polynomials, which was compatible with the accessible
computer time. The eigenvalue equation (\ref{seq}) was transformed
into $49\times 49$ matrix equation and solved in a standard
way. In some cases in order to check sensitivity of the procedure
also other sets have been used, i.e. $5\times9$ instead of $7\times7$
and consistency of results was confirmed.

One important remark should be done here. The basis we have used is
not large enough to analyze highly excited states with quantum numbers
much more than unity because of loss of the accuracy. But in this region
powerful quasiclassical methods may be applied in order to get the
solution. Indeed, one of the main questions we are going to solve
is to compare the spectra of light--cone Hamiltonian
and center of mass one for the same dynamical problem: two
massive spinless quarks connected by the straight--line string.
In 1+1 case it is known for a long time \cite{kalash}, that light--cone
and center of mass Hamiltonians for this problem are quasiclassically
equivalent. It is reasonable to assume the quasiclassical equivalence
in 3+1 case too. Therefore the main interest should be concentrated on
the lowest states and the numerical procedure we have used is mostly
adequate namely for the lowest states.

The resulting eigenvalues $M^2(\kappa, y, \alpha )$ were ordered in
their magnitude and for each state separately the minimizing value
of $y, \alpha $ was found.
The orthogonality of the resulting wave functions for minimized
parameters was then checked and appeared to be around few percents
or better for all states except two. The latter belonged to almost
degenerate states and the orthogonalization procedure for them is
made easily.

The two--dimensional wave functions have been computed for minimized
eigenvalues and formfactors and structure functions are computed
according to the formulae of Section~4.

\section{Discussion of results}

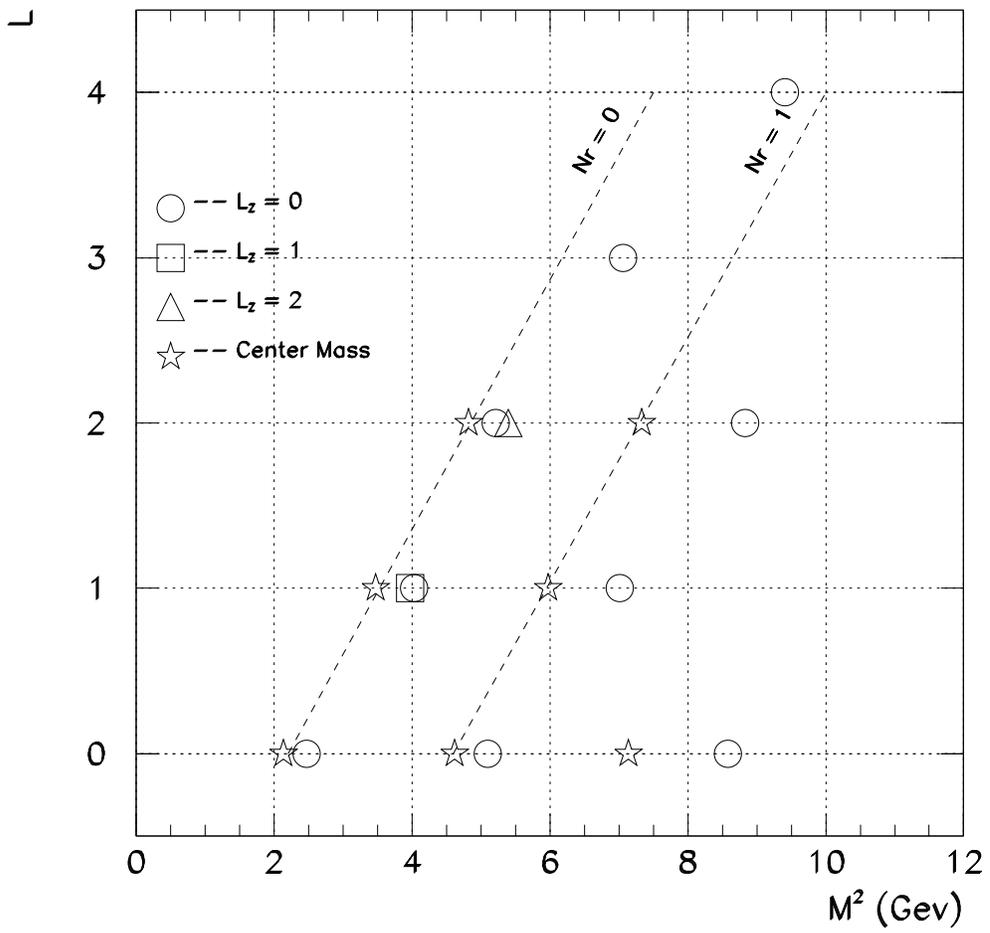
\begin{figure}
\caption{
The Chew-Frautschi plot with masses computed via the light--cone
(circles, squares and triangle for $L_{z} = 0,1,2$ respectively) and
the c.m. Hamiltonian (stars). The systematic overall mass shift is
seen as a divergence of straight lines passing through circles and stars.
The states with high $L$ or high $N_r$ (daughter trajectories) are
numerically less reliable and not shown. }
\label{fig1}
\end{figure}

We have chosen six sets of quark mass parameters, including 4 sets
of equal masses and 2 sets of unequal masses as shown in Table~\ref{tab1}.
(all masses are in units of GeV). The values of parameters
$y, \alpha, \epsilon$
are obtained by minimization procedure as explained in the previous section.
The eigenstates obtained by our numerical procedure, explained in Section~6,
are listed in Table~\ref{tab2} and shown as Fig.~\ref{fig1}.

Let us first discuss the light--cone spectrum. The Hamiltonian (\ref{ham})
depends explicitly on $L_z$, whereas the Lorentz--invariance requires
that masses do not depend on $L_z$, but depend on $L$. Consequently one
expects
that the lowest state for $L_z = 0$ corresponds to $L=0$, while the next
state with $L_z = 0$ corresponds to $L=1$ and must be degenerate in mass
with the state $L_z = 1 ,  L=1$. This is clearly seen in Table~\ref{tab2} and 
Fig.\ref{fig1}.
The next triplet of states with $L_z = 0,1,2$ should correspond to $L=2$
Two of these states are found by us and minimized, while the third
is not minimized since the procedure for this state was extremely
time--consuming. As a result two states for $L=2$ listed in Table~\ref{tab2} and
Fig.~\ref{fig1} are degenerated within 2\% accuracy, while the third state is
some 12\% up (but should come down after minimization).
The same feature is roughly present for $L=3,4$ states where we have
only one minimized state for each $L$.

At this point one should take into account another degeneration -- the dynamical
one, and to this end discuss first center--of--mass spectrum, obtained from the
c.m. Hamiltonian \cite{dub2} or by numerical diagonalization of the string equation
of motion \cite{cours}. We have computed the c.m. spectrum using the
routine from \cite{bakk}. The c.m. Hamiltonian for $L=0$ reduces to
the so--called spinless Salpeter equation \cite{lucha} which was actually
solved by us. The approximate form of the spectrum which can be easily
computed also  by the WKB method \cite{cea} can be represented as:
\be
\left( M^{(0)}(L, N_r)\right)^2 \cong 2\pi \sigma \left( 2N_r + \frac{4}{\pi}
L + \frac32 \right)
\label{nhn}
\ee
For  $L>0$ one should take into account the string contribution, absent
in spinless Salpeter equation but present in the Hamiltonian \cite{dub1}
and in \cite{cours}. This correction found in \cite{cours} is:
\be
\Delta M = -\frac{16}{3} {\sigma}^2 \frac{L(L+1)}{{(M^{(0)})}^3}
\label{dM}
\ee
The corrected masses, $M(L, N_r) = M^{(0)} + \Delta M $ are shown in 
Table~\ref{tab2}
and Fig.\ref{fig1}. They are very close to the values computed in \cite{cours} .
Now one can see that there is an approximate degeneration in mass
$M = M^{(0)} + \Delta M$
of states when one replaces one unit of $N_r$ by two units of $L$.
This is seen in Table~\ref{tab2} and even better in \cite{sim2}. The same type
of approximate mass degeneration is seen in the light--cone
spectrum -- compare
e.g. the states with $(N_r , L, L_z) = (1,0,0)\; ; \; M_{LC} = 2.25$ and
$ (0,2,0) \; ;\; M_{LC} = 2.28$.

In Fig.\ref{fig1} this degeneration is visualized as the fact that masses 
appear on the
vertical lines. This degeneration is a dynamical one and is a characteristic
feature of nonrelativistic oscillator. In our relativistic case it reveals
a new string -- like symmetry, typical for the QCD string spectrum \cite{sim2}.

Let us now compare the light--cone and the center--of--mass spectrum.
One could expect the coincidence up to an overall shift due to the
different treatment of Z--graphs in two systems. In the c.m. Hamiltonian
these Z--graphs are presented but supposed to be unimportant on the
grounds, that
the backtracking of a quark in time necessarily brings about a folding in
the string world sheet which costs a large amount of action and is
therefore suppressed. The situation is different in the light cone --
it is general belief that Z--graphs are absent here.
One also expects that the Z--graphs (and the overall shift)
should decrease if quark masses are increasing.

The comparison of the spectra can be made from Table~\ref{tab2} and 
Fig.\ref{fig1}. One can
see indeed some overall shift down in the c.m. spectrum by some 0.1 GeV
and otherwise the masses coincide within the accuracy of computation.
This fact is highly nontrivial since two quantum Hamiltonians
(the light--cone and
center--of--mass ones) are very different, they cannot be obtained from
each other by a simple boost or other simple transformation.
The light--cone Hamiltonian is rather complicated and it took the authors
more than a year to get reasonable numerical results for it.

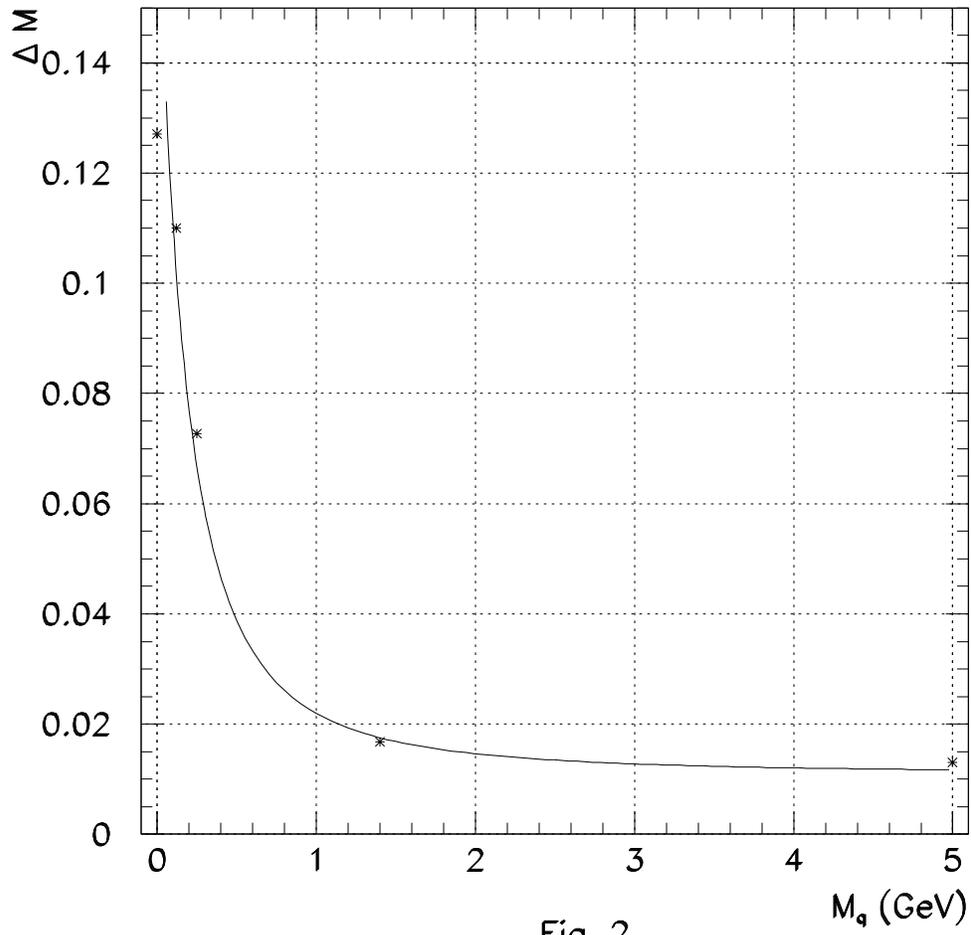
\begin{figure}
\caption{
The systematic mass shift between mass eigenvalues of the light--cone and
c.m. Hamiltonians versus quark mass. }
\label{fig2}
\end{figure}

We have also checked the quark mass dependence of the overall shift of
spectra and proved that it drops sharply with the quark mass increasing
see Fig.~\ref{fig2},
supporting the idea, that the shift is due to different treatment of
Z graphs (or self energy graphs). This fact also confirms, that the shift
is not a consequence of some systematic errors of our procedure.
In that case one should expect the decreasing of relative mass difference,
i.e. $\delta M / M$ with the quark mass increasing and not the decreasing
of absolute difference $\delta M$, which has been actually observed.

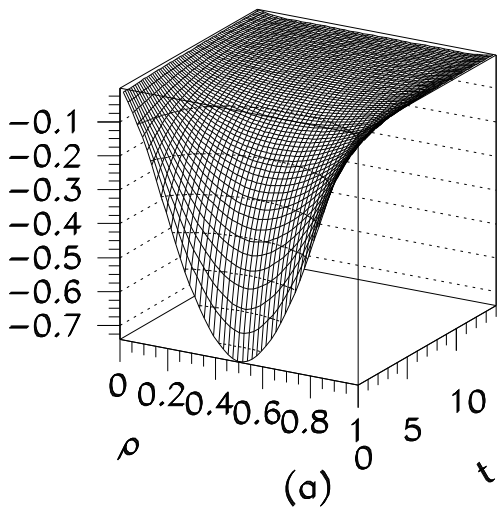
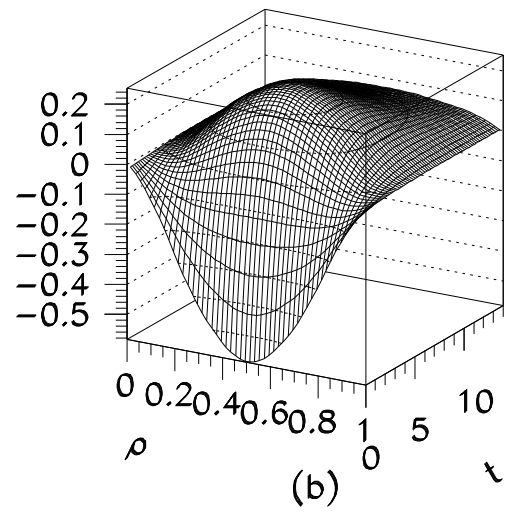
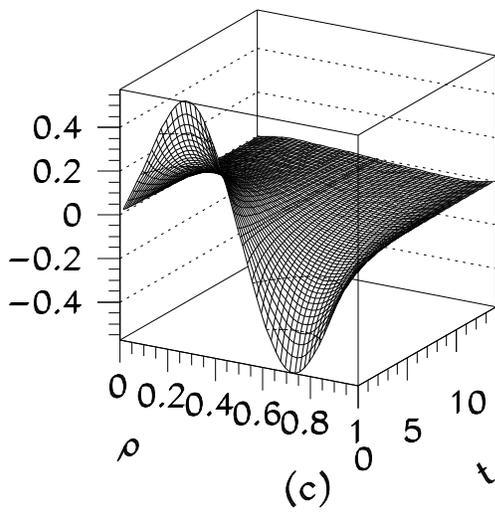
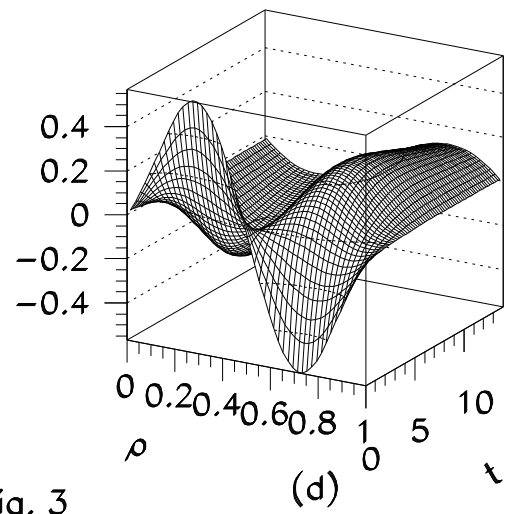
\begin{figure}
\caption{
The 3d plots of wave functions of the four lowest states of light--cone
Hamiltonian for zero quark masses. Coordinates on horizontal plane are
$0 \leq \rho \leq 1$ , $0 \leq t \leq 15$ (in units of $\sigma^{-1}$). }
\label{fig3}
\end{figure}

We now turn to eigenfunctions. One expects in this case two types of excitation:
the radial one leading to new nodes on $\rho$ coordinate and similar to
the 1+1 excited states and the $r_{\bot}$--excitation which causes  nodes in
the $t$--coordinate and associated with the orbital excitation.
This is clearly seen in Fig.~\ref{fig3}, where (a) refers to the ground 
state, (b) --
to the orbital and (c) -- to the radial excitation. A more complicated example,
combining both types of excitation is given in Fig.~\ref{fig3} (d).

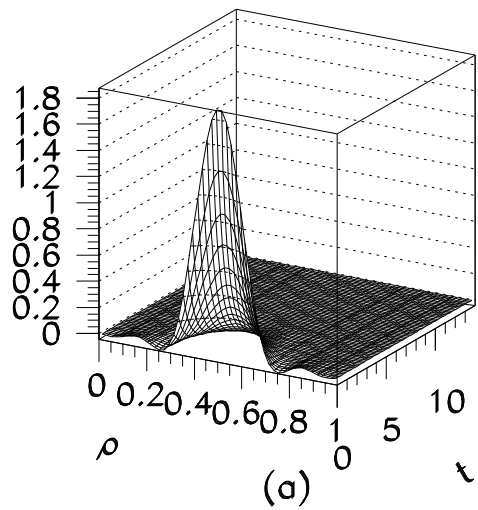
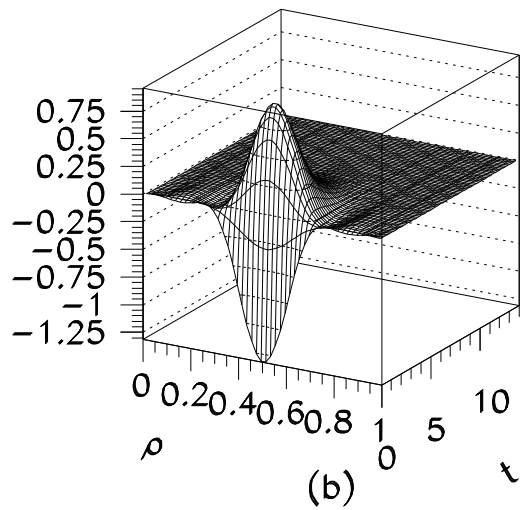
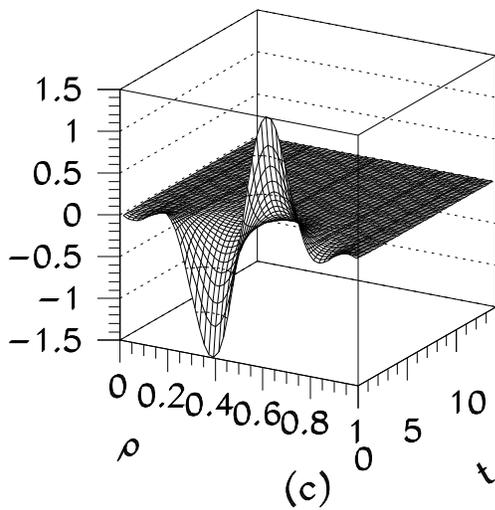
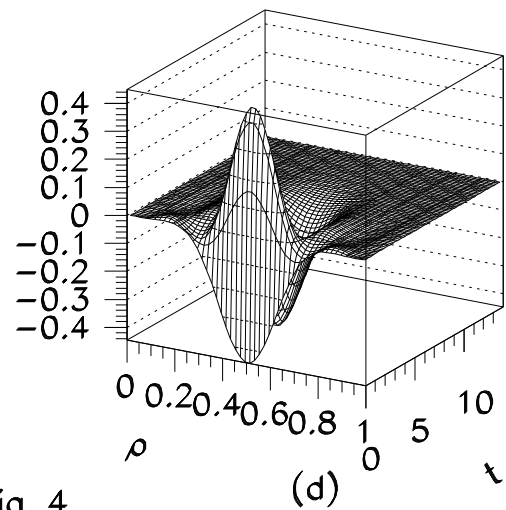
\begin{figure}
\caption{
The same as in Fig.~3 but for heavy quark masses, $m_1 = m_1 = 5$ GeV. }
\label{fig4}
\end{figure}

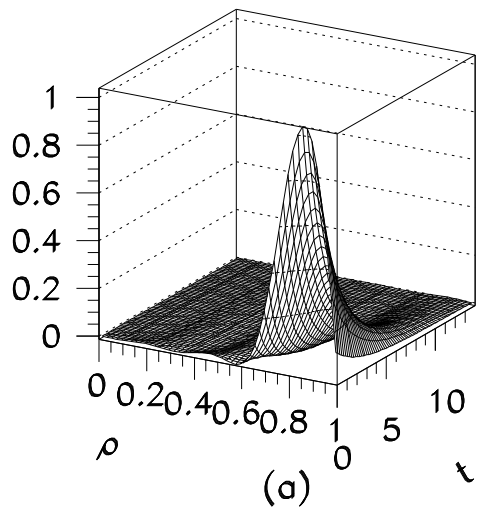
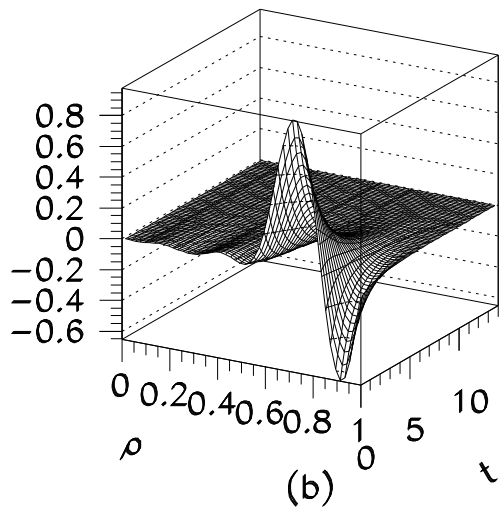
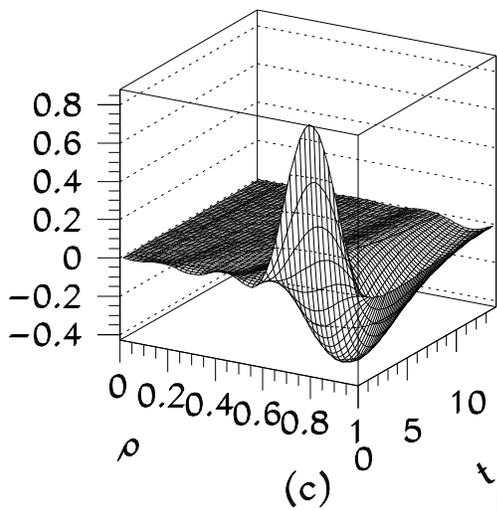
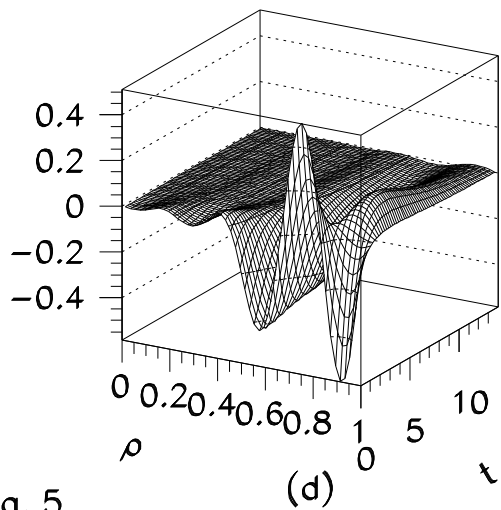
\begin{figure}
\caption{
The same as in Fig.~3 but for unequal quark masses, $m_1 = 5$, $m_2 = 0$ GeV. }
\label{fig5}
\end{figure}

As the next illustration we show in Fig.~\ref{fig4} the case of two heavy 
quarks,
demonstrating two types of excitation and also a new feature --
the actual region of parameters is squeezed to a small region near
$\rho = 1/2$. In Fig.~\ref{fig5} we demonstrate four states for the 
case of unequal
masses -- the "physical region" is shifted to one of the ends of
the [0,1] interval.

At this point it is important to find connection of our light--cone wave
function to the nonrelativistic one, usually defined in the c.m..
In \cite{dub2} it was demonstrated that this connection can be established
only when both quarks are heavy, $m_1, m_2 \gg \sqrt{\sigma} $.
In this case $\tilde y \ll 1$ (as can be found by direct minimization
of the Hamiltonian (\ref{ham})) -- the string transforms into the potential
and loses its material and momentum contents. One can introduce as in
\cite{dub2} the relative momentum $p_z$ and relative coordinate $r_z$:

\be
p_z = (m_1 +m_2) \left( x - \frac{m_1}{m_1 + m_2} \right)
\label{pz}
\ee
\be
r_z = \frac{P_+ r_-}{m_1 + m_2}
\label{io}
\ee
and the $M$ operator can be written as:
\be
 M \approx m_1 + m_2 + \frac{1}{2 \tilde{m}}
({\vec{p}}~^{2}_{\bot} + {\vec{p}}~^{2}_{z}) +
\sigma \sqrt{r^{2}_{\bot} + r^{2}_{z}}
\label{Mapp}
\ee

Hence one can write the momentum--space nonrelativistic wave function
$\Psi({\vec p}~^{2})$ directly through the light--cone variables:
\be
\Psi({\vec{p}}~^{2}) = \Psi \left[p_{\bot}^2 + (m_1 + m_2)^2\> \left(
 x - \frac{m_1}{m_1 + m_2} \right)^2 \right]
\label{psip2}
\ee

This representation is valid in the large mass limit $m_i \gg \sqrt{\sigma}$
stated above and in addition near the center of the $x$--distribution,
i.e. when $|x - m_1 / (m_1 + m_2)| \ll 1$. The width of the peak in $x$
variable is proportional to $(m_1 + m_2)^{-2}$ and is very narrow for
heavy quarks. The form (\ref{pz}) is not correct for $x$ at the ends of the
interval, i.e. $x = 0,1$ (remember that for large $m_i$ the extremum value
of $\tilde{y}$ tends to zero and $x = \rho$ changes in the interval
[0,1]). Indeed the exact wave function as discussed in Section~4 vanishes
linearly at $x = 0,1$, while the r.h.s of (\ref{psip2}) stays nonzero.
Moreover, the Jacobean {$\cal{J}$} of the phase space
$d^3 p = {\cal{J}} dx d^2 p_{\bot}$ is constant ${\cal{J}} = m_1 + m_2$
and does not change this conclusion.

The correspondence of the c.m. and light--cone wave functions is lost
if quark masses $m_i$ are of the order of $\sqrt{\sigma}$ or less.
The physical reason is that the role of dynamics is now 100\%
important and the dynamics is {\underline {different}} in different
frames: the light--cone Hamiltonian and wave functions are connected
with those of the center--of--mass by a dynamical transformation,
which includes {\underline {nonkinematical}} Poisson operators.
Hence one cannot hope to obtain one wave function from another
using only transformation including the kinematical (free) part of the
operator $M^2$. Moreover, a glance at the operator $M^2$ in (\ref{ham})
helps to realize that the separation of $M^2$ into the purely
kinetic part (containing only momentum and masses) and purely
potential one (containing only relative coordinates) is
{\underline {impossible}} at all: e.g. the string enters through
the term $(P_+r_-)^2$ and mixed terms like $p_{\bot}r_{\bot}$
are present everywhere. This circumstance limits the
use of simple recipes known in literature \cite{chung},
which connect the c.m. and light--cone wave functions.
In particular, the ansatz for $p_z$ suggested in \cite{chung}
and used for equal quark masses
\be
p_z = \sqrt{\frac{m^2 + k_{\bot}^2}{x(1-x)}} \left( x - \frac12 \right)
\label{g}
\ee
coincides with our form (\ref{io}) rigorously derived in \cite{dub1}
only for large $m$, $m \gg \sqrt{\sigma} $ and $x$ in the narrow region
around $x = 1/2$ but differs at the tails of the wave function.
For light quark masses, $m \le \sigma$ and small $p_{\bot}$
the relation (\ref{g}) yields incorrect
results which can be seen in the almost constant behaviour in
the interval [0,1] of the light--cone wave function, produced by the
insertion of (\ref{g}) into the c.m. wave function, typically
$$
\psi({\vec p}~^{2}) \sim \exp (- a^2 {\vec p}~^{2}) 
\to \exp (- a^2 [ p_{\bot}^2 + \frac{m^2 + p_{\bot}^2}{x(1-x)}
(x-\frac12)^2 ] )
$$
is insensitive to $x$ for $m,p_{\bot} \to 0$, whereas
the exact light--cone wave function is decreasing as $x(1-x)$, see 
Fig.~\ref{fig6}.

At the same time for heavy quark masses, $m \gg \sqrt{\sigma}$ the two
functions, one transformed by (\ref{g}) from the c.m. and another is
a genuine light--cone wave--function solution of (\ref{seq}), are very
close to each other. This can be seen in Fig.~\ref{fig7} for $m = 1.4$ GeV.

We now turn to the formfactor of computed states 1-4, see Table~\ref{tab1},
presented in Fig.~\ref{fig6}. The main feature of the Fig.~\ref{fig6} 
is a very slow
decrease of $F(q^2)$ with $q^2$ which signals in particular a small
radius of states. Indeed the values of $\sqrt{<r^2>}$
are too low ($\sim 0.338$ $fm$ for massless quarks). This fact is
in agreement with the earlier c.m. calculations of \cite{carl}.

Note that color Coulomb and spin interaction can only decrease
radius. In the same Fig.~\ref{fig8} are also shown formfactor calculated
in the c.m. system for the equal mass values listed in Table~\ref{tab1}
(cases 1--4). These formfactors can be calculated either in terms
of light--cone variables as in \cite{chung} and \cite{drell}, or
simply using the nonrelativistic expression
\be
  F(q^{2}) = \int | \psi_{cm}(r)|^{2} \frac{sin(q r /2)}{q r/2} d^{3}r
\label{Fq2}
\ee
since both integrals can be transformed one into another by a change
of variables. One can see in Fig.~\ref{fig8}, that the c.m. formfactor is
systematically below that of the light--cone.

Finally we turn to the quark--distribution function $q(\rho)$.
It is computed through the light--cone wave--function $\psi(\rho ,t)$
using (\ref{qr}) and shown in Fig.~\ref{fig9}. One can see the symmetric 
behaviour
of $q(\rho)$ with respect to reflection $\rho \to 1-\rho$. At the ends
of the interval $q(\rho)$ vanishes like $\rho^{2}$ and $(1-\rho)^2$, in
the agreement with the $1/q^3$ behaviour of the formfactor at large
$q$ due to the Drell-Yan-West relations \cite{west}.

Note the narrowing of the peak in $q(\rho)$ in Fig.~\ref{fig9} for increasing
quark masses.

\section{Conclusion}

The present paper is the first in the planned series of papers devoted
to the systematic study of nonperturbative contribution to formfactors,
quark distributions and high-energy scattering amplitudes.

The main physical idea of our approach is that the most part of
nonperturbative dynamics in QCD is due to the QCD string, and the latter
is described by the Nambu-Goto part of the Hamiltonian, which was written
before in the c.m. \cite{dub2} as well as in the light-cone coordinates
\cite{dub1}.

Only valence part of the Fock's column was considered above in the paper,
also for simplicity spins and perturbative gluon exchanges are neglected.
To do the systematic comparison with experiment all these three
simplifications should be eliminated. Let us discuss their effect point
by point. The higher Fock states are necessary to reproduce the Regge
behaviour of $q(x) \sim x^{-\alpha_{\rho}(0)}$
at small $x$ (and at $x \to 1$ for quark distributions of hadrons in
high-energy scattering). Here comes the first crucial point; to be answered
in the second paper of this series, what is the QCD reggeon?

In our method the higher Fock states, constituting the QCD reggeon,
correspond to several gluons propagating in the nonperturbative
background and therefore confined to the excited Nambu-Goto surface \cite{sim3}.
These states are in one-to-one correspondence with the excited Nambu-Goto
string states. This is the picture at large distances; at small distances
smaller than the vacuum correlation length (width of the Nambu-Goto string)
$T_{g} \sim 0.2$ $fm$ \cite{debb},  the string disappears and the usual
perturbative gluon exchanges reappear.

The effect of spin of light quarks is highly nontrivial \cite{sim1}.
It leads to the creation of the new vertex, which yields the constituent
quark structure. Physically one may imagine this structure as being due
to the light--quark walks around the end of the string.

Having said all this, what is the lesson of the present work and of it
possible development?

The first lesson is that valence quark component
can be successfully dynamically computed on the light cone; the string on
the light cone is physically and mathematically well defined. The spectrum
obtained on the light cone for the first time reasonably coincides with that
of the c.m. Hamiltonian for the string with quarks. Moreover, the nonperturbative
wave function obtained directly on the light cone allows to calculate
nonperturbative contributions to the formfactor and structure function.

The second lesson is that the formfactor computed directly on the light cone
is close to of c.m. for small $q$ (see Fig.~\ref{fig7}), but is systematically
above the c.m. formfactor for larger $q$. This is not surprising, since in the
light--cone formfactor there is a mechanism of the "redistribution" of the
momentum $q$ between the quarks, since $q$ enters the light--cone formfactor
(\ref{Ff1}) multiplied with $(1 - x)$, so that the larger $q$, the smaller
is $(1 - x)$ and the wave function does not decrease too fast.
Physically it means that at large $q$ the configuration survives where
the spectator quark gets as little momentum as possible so that it can
be easily turned together with the active quark. This is exactly what
is called the Feynman mechanism \cite{feynm}.

The third lesson is that the minimal string plays only a passive role on
the light cone, namely it participates in sharing of the total momentum
and carries the part equal to $<\tilde{y}>$, but it does not produce the
$x$--distribution in structure function,
which could simulate the gluonic structure
function. The reason is that the string variable -- the einbein field
$\nu$ -- is quasiclassical and has no dispersion.

The value of $<\tilde{y}>$ computed according to (\ref{ytil}) depends
on quark masses and is listed in Table.~\ref{tab1}. It is resonable that
$<\tilde{y}>= 0.22$ is smaller than the experimental value of overall
gluon momentum, $0.55$, since in our picture the difference should
be filled in by higher Fock components.

The fourth lesson comes from the comparison of the computed quark
distribution, Fig.~\ref{fig9}, with the experimental data for the pionic
structure function \cite{sutt}. Behaviour of $q(\rho)$ at small $\rho$ and
small $(1- \rho)$ is symmetric in Fig.~\ref{fig9}, while in reality 
$q(x)$ should
rise at small $x$ like $x^{-\alpha_{\rho}(0)} \sim x^{-0.5}$
(we neglect at this point the difference
between $x$ and $\rho$, which is due to $<\tilde{y}>$). This peak at the small
$x$ should be filled in by the contribution of higher Fock components,
containing additional gluons on the string, as was discussed above. The
behaviour of $q(\rho)$ at $\rho = 1$, which is calculated to be
$(1-\rho)^{2}$ will be also changed into $(1 - \rho)$ due to gluon
exchanges, which account for the formfactor asymptotic $1/q^{2}$ at large
$q$, and the Drell-Yan-West duality ensure the $(1 - \rho)$ behaviour
around $\rho = 1$.

Finally, the formfactor calculated above in the paper, Fig.~\ref{fig8} shows too
little radius of the "pion" $<r^2> \approx (0.34 \; fm)^2$ as compared
with the experimental one $<r^2> \simeq (0.67 \; fm)^2$. This fact is
in qualitative agreement with other calculations, where the
c.m. wave--function was used \cite{carl}, and some authors assumed
as in \cite{altar2} that quarks have "internal" structure and their
own radius which should be added to the "body radius" to reproduce
the experimental value. This fact of small body radius seems to
be a necessary consequence of the simple string + point-like quarks
picture, and probably cannot be cured by the higher Fock components.

The calculation of the quark structure as produced by the spin and chiral
effects is thus an interesting and fundamental problem which will be
discussed in another paper of this series.

\acknowledgements
The calculations of the c.m. wave--function have been done using
the codes by B.L.G.~Bakker, the Free University of Amsterdam. The
authors are very grateful to him for the submitting of his codes and
valuable explanations.

The authors are grateful to A.B.~Kaidalov for numerous discussion
of QCD reggeon picture, and to Yu.S.~Kalashnikova and A.~Nefediev
for discussions concerning different questions of hamiltonian
dynamics and to V.A.Novikov for useful comments.

This work was supported in part by the RFFI grant N 96--02--19184a and
95-048-08.
Yu.S. was supported in part by the INTAS grant 93-79.

\appendix
\section{}


This appendix is based on the material of Ref.\cite{dub1}

Given a $q\bar{q}$ Green's function in the coordinate space
$G(x\bar{x};y\bar{y})$, where $x\bar{x}(y\bar{y})$ are final (initial)
4--coordinates of quark and antiquark, one can define the Hamiltonian $H$
through the equation (in the euclidean space--time)
$$
\frac{\partial G}{\partial T} = - HG
\eqno(\mbox A.1)
$$
where $T$ is an evolution parameter corresponding to some choice of a
hypersurface $\Sigma$. In a particular case of the c.m. Hamiltonian the
role of $T$ is played by the center--of--mass euclidean time coordinate
$T=(x_4+\bar{x}_4)/2$ and the hypersurface $\sum$ is a hyperplane
$x_4=\bar{x}_4=const.$

With the notations  for the vectors $a_{\mu},b_{\mu}$
$$
ab= a_{\mu}b_{\mu} = a_ib_i-a_0b_0=a_{\bot}b_{\bot}+a_+b_-+a_-b_+,
$$
$$a_{\pm}=\frac{a_3\pm a_0}{\sqrt{2}},$$
one can define the hypersurface $\sum$ through the $q\bar{q}$ coordinates
$z_{\mu}.\bar{z}_{\mu}$ as
$$
z_+(\tau)=\bar{z}_+(\bar{\tau})
$$
and the kinetic part of the action $A$
$$
A=K+\bar{K}+\sigma S_{min},
$$
has the form
$$
K+\bar{K} = \frac{1}{4} \int^s_0 \dot{z}^2_{\mu}(\tau)d\tau+
$$
$$
+\frac{1}{4}
 \int^{\bar{s}}_0 \dot{\bar{z}}^2_{\mu}(\bar{\tau})d\bar{\tau}+
  \int^{s}_0m_1^2 d\tau+ \int^{\bar{s}}_0m_2^2 d\tau =
$$
$$=
\int^T_0dz_+\left[ \frac{\mu_1}{2}(\dot{z}^2_{\bot}+2\dot{z}_-)+\right.
$$
$$
\left. +\frac{\mu_2}{2}
(\dot{\bar{z}}^2_{\bot}+2\dot{\bar{z}}_-)
+\frac{m_1^2}{2\mu_1}+\frac{m^2_2}{2\mu_2}\right]
$$
  where we have defined
$$
  2\mu_1(z_+)=\frac{\partial z_+}{\partial \tau}~;~~
  2\mu_2(z_+)=\frac{\partial \bar{z}_+}{\partial \tau}
$$

The minimal surface $S_{min}$ is formed by connecting $z_{\mu}(z_+)$ and
$\bar{z}_{\mu}(z_+)$ with the same value of the evolution parameter $z_+$, i.e.
$$
  S_{min}= \sigma \int^T_0 dz_+\int^1_0 d\beta [\dot{w}^2
  w'^2-(\dot{w}w')^2]^{1/2}
$$
  where
$$
  w_{\mu}(z_+; \beta) = z_{\mu}(z_+) \beta + \bar{z}_{\mu}(z_+) (1-\beta)
$$
  and dot and prime denote derivatives  in  $z_+$ and $\beta$ respectively
  throughout this Appendix.

  We now introduce "center--of--mass" and relative coordinates,
$$
  \dot{R}_{\mu}=x\dot{z}_{\mu} + (1-x)\dot{\bar{z}}_{\mu}~,~~
  \dot{r}_{\mu}=\dot{z}_{\mu}-\dot{\bar{z}}_{\mu}
$$
where the variable $x$ is defined
 from the requirement that the term
$\dot{r}_{\bot}\dot{R}_{\bot}$ should be absent in the action.
This yields:
$$
x=\frac{\mu_1+\int \nu\beta d\beta}{\mu_1 + \mu_2 + \int \nu d\beta}
$$
Then for the Green function one obtains:
$$
G(x\bar{x};y\bar{y})= \int D{\mu}_1(z_+) \; D\mu_2(z_+)D\nu \;
DR_{\mu} \; Dr_{\mu} \; e^{-A}
$$
where the action $A$
$$
A=\frac{1}{2}\int dz_+\left\{\frac{m^2_1}{\mu_1}+\frac{m^2_2}{\mu_2}+
a_1(\dot{R}^2_{\bot}+2\dot{R}_-)+a_3\dot{r}^2_{\bot}+ \right.
$$
$$
+ \int\frac{\sigma^2}{\nu}d\beta \; r^2_{\bot} -
\frac{(r_-+\dot{R}_{\bot}r_{\bot}+(<\beta>-x)\dot{r}_{\bot}r_{\bot}
)^2}{r^2_{\bot}(\int \nu d\beta)^{-1}}-
$$
$$
\left. - \frac{(\dot{r}_{\bot}r_{\bot})^2\int\nu(\beta-<\beta>)^2d\beta}
{r^2_{\bot}}
\right\},
\eqno(\mbox A. )
$$
The following notation was used:
$$
a_1=\mu_1+\mu_2+\int^1_0\nu(\beta)d\beta
$$
$$
a_3=\mu_1(1-x)^2+\mu_2x^2+\int^1_0 \nu(\beta)(\beta-x)^2 d\beta
$$
Integration over $DR_{\mu}$ leads to an important constraint:
$$
a_1 = P_+
$$
Furthermore we go over into the minkowskian space, which means that
$$\mu_i\to - i\mu^M_i~,~~\nu \to - i\nu^M$$
$$
a_i\to - ia^M_i, ~~A\to - i A^M
 $$
 For the minkowskian action we obtain  (omitting from now on the superscript
 $M$ everywhere)
 $$
 A^M= \frac{1}{2} \int dz_+\left\{
 -\frac{m^2_1}{\mu_1}-\frac{m^2_2}{\mu_2}+a_3\dot{r}^2_{\bot}- \int
 \frac{\sigma^2 d\beta }{\nu}r^2_{\bot}-\right.
$$
$$
  -\left.\nu_2\frac{(\dot{r}_{\bot}r_{\bot})^2}{r^2_{\bot}}-
 \frac{\nu_0a_1}{(\mu_1+\mu_2)r^2_{\bot}}[r_-
+(<\beta>-x)\dot{r}_{\bot}r_{\bot}]^2 \right\}
 $$
 and using (4) one easy obtains the Hamiltonian (\ref{ham})
 which opened the main text.

To complete the Hamiltonian formulation of our problem we define canonical
momenta for the coordinates $\dot{R}_-$ and $\dot{r}_-$.
As it was shown, canonically conjugated momentum to the
$\dot{R}_-$ is $P_+=a_1$. Situation with $\dot{r}_-$ is more subtle.
In order to clarify the situation let us start with the general
form of the $q\bar{q}$ Green's function in the
Feynman--Schwinger formalism
$$
G(x,y)=\int ds~d\bar{s}DzD\bar{z}`^{-K-\bar{K}}<W(C)>
$$
to impose boundary conditions, one can rewrite $DzD\bar{z}$ using
discretization
$$ \xi_n\equiv z(n)-z(n-1), \bar{\xi}_n=\bar{z}(n)-\bar{z}(n-1)$$
$$
DzD\bar{z} = \prod_{n,n'}d\xi_nd\bar{\xi}_{n'} \;  dp~dp' \times
$$
$$
\times exp \left\{ 
ip(\sum\xi_n + X-Y) + ip'(\sum\bar{\xi}_{n'}+X-Y) \right\}
$$
One can introduce the total and relative momenta
$$
P=p+p',~~q=\frac{p-p'}{2};
$$
and $\dot{R}=\Delta R_n /\varepsilon; ~~N\varepsilon
=T, $  one has
 $$
 \xi_nx_n+\bar{\xi}_n(1-x_n)=\Delta R_n
 $$
$$
 \xi_n-\bar{\xi}_n=\Delta r_n
 $$
Expressing $\xi_n,\bar{\xi}_n$ through $\Delta R_n, \Delta r_n$
 and going over to the momentum representation of $G$ one obtains
 $$
 G(P)=\int dq \prod_{n,n'} d\Delta R_n d\Delta r_{n'} \times 
$$
$$
 \times exp \left\{ iA + iP\sum_{n}\Delta
 R_n+i\sum_n(\frac{1}{2}P(1-2x)+q)\Delta r_n \right\}
$$
 where $A=\int^T_0d\tau{\cal L} = \int^T_0 d\tau (K+\bar{K}-\sigma
 S_{min})$
 One can now introduce the Hamiltonian form of the path integral via
 $$
 \int Dx e^{i\int {\cal L}d\tau }= \int Dx Dpe^{ip_k\dot{x}_k-i\int {\cal
 H}d\tau}
 $$
 and rewrite the first two exponents as
 $$
 exp \left\{ i\int P_i \dot{R}_i d\tau + i \int[\frac{1}{2} P_i
 (1-2x(\tau))+q_i]\dot{r}_id\tau \right\}
$$
 The term proportional $q_i$ disappears because of boundary conditions
 $r_{\mu}(0) = r_{\mu}(T) =0$, and one obtains
 $$
 p_+=\frac{1}{2} P_+(1-2x),~~[p_+,r_-]=-i
 $$
 One can rewrite this in the form
 $$
 P_+r_-=i\frac{\partial}{\partial x}, ~~[P_+r_-, x]=i
 $$

\section{}

The aim of this appendix is to derive the Eq.(\ref{seq}) from Eq.(\ref{ham}).
The first thing to do is to express the Hamiltonian (\ref{ham}) as :
$$
H=\frac{M^2}{2 P_+}
\eqno (\mbox{B}.1)
$$
Then using definition (\ref{x}) we obtain the following form for $M^2$:
$$
 M^2= \left\{ \frac{m^2_1}{x-{\tilde y}_0 <\beta>}+
\frac{m^2_2}{1-x - {\tilde y}_0 (1-<\beta>)}+
 \frac{L_z^2}{{\tilde c}\>r^2_{\bot}} + \right.
$$
$$
 + \left[ \frac{1}{x-{\tilde y}_0 <\beta>} + \frac{1}{1-x - {\tilde y}_0
(1-<\beta>)} \right] \times
$$
$$
\times \frac{[p_{\bot}r_{\bot}+ \tilde {\gamma} (P_+r_-)]^2}{ r^2_{\bot}} +
$$
$$
+ \left.\int \frac{\sigma^2}{\tilde y} r^2_{\bot} d\beta +
\frac{{\tilde y}_0}{1-{\tilde y}_0} \frac{(P_+r_-)^2}{r^2_{\bot}}  \right\}
\eqno (\mbox{B}.2)
$$
here
$$
{\tilde c} =
(x-{\tilde y}_0<\beta>)(1-x)^2 + 
$$
$$
[1-x -{\tilde y}_0 (1-<\beta>)] x^2 + \int_0^1\tilde y (\beta - x)^2 d\beta
$$
and
$$
\tilde \gamma = \frac{{\tilde y}_0}{1-{\tilde y}_0}\> (<\beta> - x)
$$
As it was mentioned above, it is more convenient to express $M^2$ through
a new variable ${\rho}$:
$$
\rho = \frac{1}{
1 - {\tilde y}_0} (x - {\tilde y}_0 <\beta>)
$$
Substituting this variable into (B.2) one obtains:
$$
 M^2= \frac{1}{1-{\tilde y}_0} \left\{ \frac{m^2_1}{\rho}+
\frac{m^2_2}{1-\rho}+
\frac{L_z^2}{{\tilde c}\>t} + \right.
$$
$$
+ \left( \frac{1}{\rho} + \frac{1}{1-\rho} \right) \times
\frac{[p_{\bot}r_{\bot}+ \tilde {\gamma} (P_+ r_-)]^2}{t} +
$$
$$
+ \left.(1-{\tilde y}_0) \int \frac{\sigma^2}{\tilde y} t d\beta +
\frac{{\tilde y}_0 (P_+r_-)^2}{t}  \right\}
\eqno (\mbox{B}.3)
$$
where the notation $t = r_{\bot}^2$ was used.

Quantization of the above expression  is done according to the canonical
commutation relations:
$$
\{p_{\bot}^{k}, r_{\bot}^{j}\} = -i {\delta}^{kj}
$$
$$
\{x, (P_+r_-)\} = -i
$$
We are looking for the wave function of the problem given in the mixed
coordinate -- momentum representation $\psi = \psi(\rho, t)$, so one has
to substitute into (B.3) the operators:
$$
(P_+r_-) = i\>\left(\frac{1}{ 1-{\tilde y}_0}\right)
\frac{\partial }{ \partial \rho}\;
\; ;\;\; p_{\bot}^k = -i\>\frac{\partial }{ \partial r_{\bot}^k}
$$
The important point is
the operators ordering. We use the Weil ordering rule, i.e.
$$
AB \to \frac12 (\hat A \hat B + \hat B \hat A)
$$
for any noncommuting operators $A$ and $B$.
Let us notice that $\tilde y$ explicitly depends on $t$ according
to (\ref{y}) and hence should also be differentiated during the ordering
procedure.
The final result for the operator ${\hat M}^2$ may be found by the
straightforward calculations, it is
$$
{\hat M}^2 =
A_1\>\frac{{\partial}^2}{{\partial t}^2} + A_2 \>
\frac{{\partial}^2}{{\partial \rho}^2} + A_3 \>
\frac{{\partial}^2}{{\partial t}{\partial \rho}} +$$
$$
+ A_4\>\frac{\partial}{\partial t}  + A_5 \frac{\partial}{\partial \rho}
 + A_6
\eqno(\mbox{B}.4)
$$
where the coefficients $A_i$ :

$$
A_1(\rho, t) =
-\frac{4t}{\rho(1-\rho)} \;
\frac{1+\alpha t}{[1 + (\alpha - y)t]}
$$
$$
A_2(\rho, t) =
-\frac{y(1+\alpha t)}{[1+(\alpha - y)t)]^3} \>
\left[
yt\frac{(\rho - <\beta>)^2}{\rho(1 - \rho)} + 1 + \alpha t
\right]
$$
$$
A_3(\rho, t) =
-4yt
\frac{(1+\alpha t)}{[1+(\alpha - y)t]^2}
\frac{(\rho - <\beta>)}{\rho (1-\rho)}
$$
$$
A_4(\rho, t) =
-\frac{1}{\rho(1-\rho)} \times
$$
$$
 \times \left\{ 4
\frac{yt}{[1+(\alpha - y)t]^2} \> + \>
4\frac{1+\alpha t }{1+(\alpha - y)t} \right. +
$$
$$
+ \left. 2 yt \frac{(1+\alpha t)}{[1+(\alpha - y)t ]^2}
\frac{({\rho}^2 - 2\rho <\beta> + <\beta>)}{\rho(1-\rho)} \right\}
$$
$$
A_5(\rho, t) =
-2y
\frac{[1+(\alpha + y)t]}{[1+(\alpha - y)t]^3}
\frac{(\rho - <\beta>)} {\rho(1-\rho)} -
$$
$$
-y^2 t \frac{(1+\alpha t)}{[1 + (\alpha - y)t]^3} \times
$$
$$
\times \frac{(\rho - <\beta>)[(\rho(1-2<\beta>) + <\beta>]}{[\rho(1-\rho)]^2}
$$
$$
A_6(\rho, t) =
\frac{1+\alpha t }{1+(\alpha - y)t}
\left( \frac{m_1^2}{\rho} + \frac{m_2^2}{1-\rho} \right) + 
\frac{1+\alpha t}{y} +  
$$
$$
+ \frac{L_z^2}{t}\frac{(1+\alpha t)^2} {a_6} -
\frac{y^2 t \; (1+\alpha t)}{[1+(\alpha - y)t]^3} \times 
$$
$$
\times \frac{[{\rho}^3(1-2<\beta>)
- 3<\beta>^2\rho(1-\rho) + <\beta>^2]}{[\rho(1-\rho)]^3} -
$$
$$
-\frac{y}{\rho(1-\rho)}
\frac{[3- (\alpha - y)t]}{[1+(\alpha - y)t]^3} -
$$
$$
-y\frac{({\rho}^2 - 2<\beta>\rho + <\beta>)}{[\rho(1-\rho)]^2}
\frac{[1+(\alpha + y)t]}{[1+(\alpha - y)t]^3}
$$
where: 
$$
a_6 = yt[1 + (\alpha - y)t](\rho - <\beta>)^2 + 
$$
$$
+[1+(\alpha - y)t](1+\alpha t)
\rho(1-\rho) + yt(1+\alpha t)\gamma
$$

\begin{figure}
\caption{
The 3d plots of the ground--state wave functions $\Psi(\rho , t)$,
computed via the light--cone Hamiltonian (upper part) and via the c.m.
Hamiltonian, with the standard substitution (\ref{g}) (lower part) for
zero quark masses. }
\label{fig6}
\end{figure}

\begin{figure}
\caption{
The same as in Fig.6 but for heavy quark masses, $m_1 = m_1 = 1.4$ GeV. }
\label{fig7}
\end{figure}

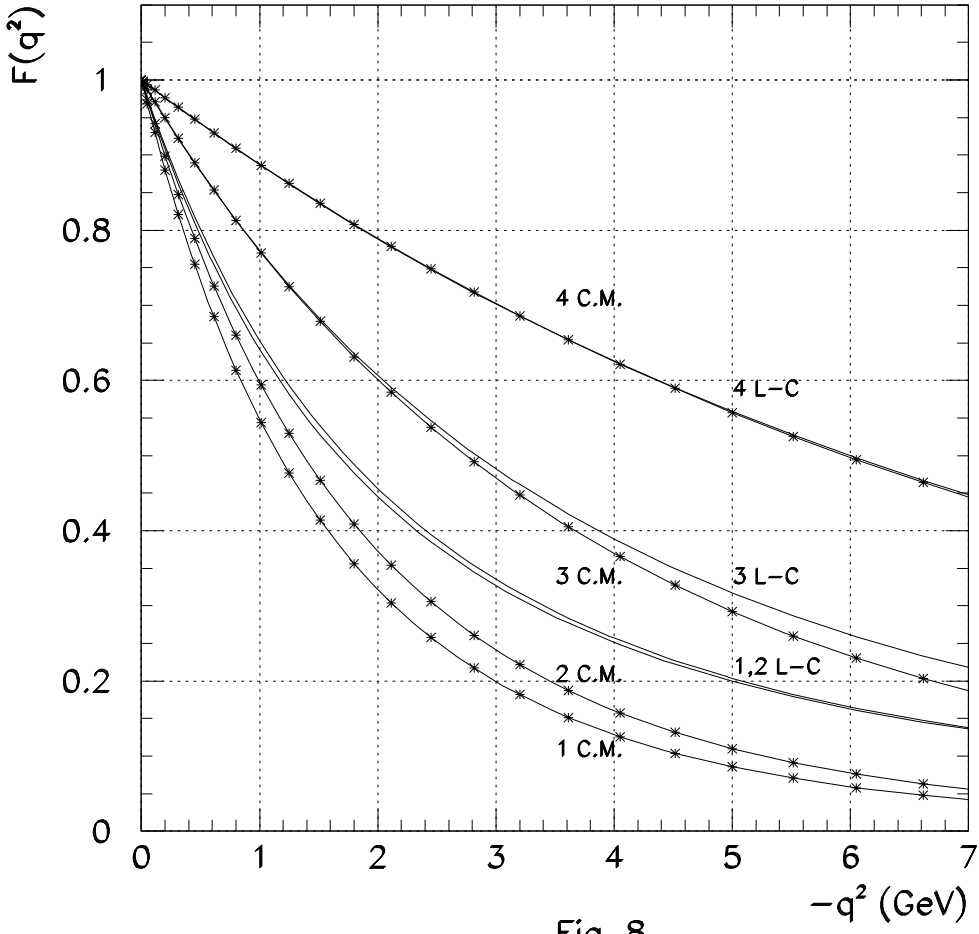
\begin{figure}
\caption{
Formfactors  calculated  with  light--cone  wave--functions  for  the  cases
1--4 of Table~\ref{tab1} (solid lines) and with the c.m. wave--functions
(lines with stars). }
\label{fig8}
\end{figure}

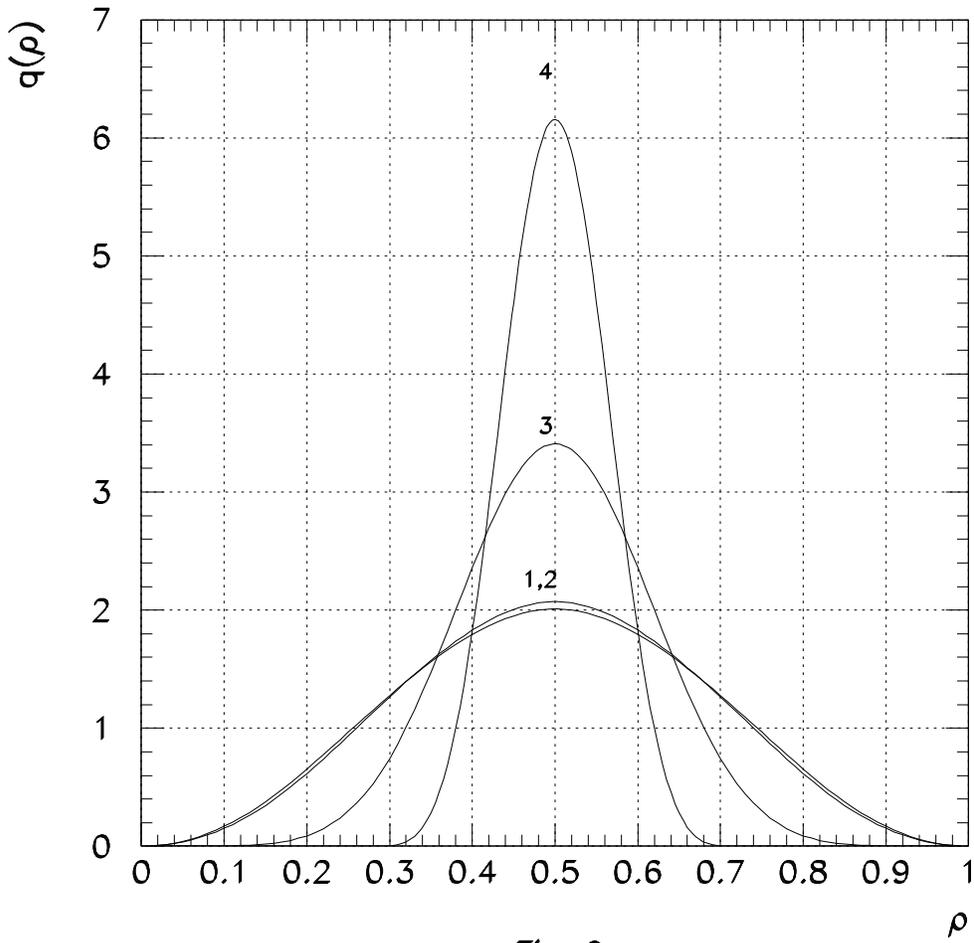
\begin{figure}
\caption{
The quark--distribution function $q(\rho)$ computed with light--cone
wave--functions for the cases 1--4 of Table~\ref{tab1}. }
\label{fig9}
\end{figure}

\begin{table}
\caption{
Quark masses and minimizing values of variational parameters for six cases,
computed in the paper together with $ < r^2 > $ and $ < \tilde{y} > $
for each case. }
\begin{tabular}{|c|c|c|c|c|c|c|c|c|}
& $m_{q_1}$ & $m_{q_2}$ & $y$ & $\alpha$ &
$\varepsilon$ & $\langle \beta \rangle$ &
$\langle \tilde{y} \rangle$ & $\sqrt{\langle r^2 \rangle} fm$  \\
\tableline
Case 1 & 0.00 & 0.00 & 0.40 & 0.75 & 1.0 & 0.50 & 0.223 & 0.338 \\
\tableline
Case 2 & 0.25 & 0.25 & 0.35 & 0.75 & 1.0 & 0.50 & 0.190 & 0.329 \\
\tableline
Case 3 & 1.40 & 1.40 & 0.20 & 0.75 & 1.7 & 0.50 & 0.081 & 0.249 \\
\tableline
Case 4 & 5.00 & 5.00 & 0.10 & 1.50 & 3.5 & 0.50 & 0.017 & 0.1675 \\
\tableline
Case 5 & 1.40 & 0.00 & 0.20 & 0.50 & 1.1 & 0.10 & ---- & ---- \\
\tableline
Case 6 & 5.00 & 0.00 & 0.07 & 0.25 & 1.5 & 0.13 & ---- & ---- \\
\end{tabular}
\label{tab1}
\end{table}

\begin{table}
\caption{
The light--cone mass eigenvalues (mass in L--C -- the fourth column),
and the c.m. mass eigenvalues without (mass in C--M -- the fifth column)
and with string correction ($M^{(0)} - \Delta M$ in CM -- the sixth column)
The first 3 columns contain the quantum numbers assignment for the given
state, and the last two columns -- the minimizing values of variational
parameters $y$ and $\alpha$. All eigenvalues are computed for quark masses
equal to 0.12 GeV. The masses marked by asterix are minimized with parametres
$y$, $\alpha$ listed in the last two columns.  }
\begin{tabular}{|c|c|c|l|c|c|c|c|}
$N_r \rule{0pt}{0.5cm}$ & $L$ & $L_z$
& Mass & $M^{(0)}$ & $M^{(0)} - \Delta M$ & $ y $ & $\alpha$ \\
& & & in L--C &in C--M& in C--M  & & \\
\tableline
  0 & 0 &  0 & 1.5704 * & 1.4604 & 1.4604 &  0.40 & 0.80 \\
&&&&&&& \\
  0 & 1 &  0 & 2.0063 * & 1.9236 & 1.8637 &  0.15 & 0.20 \\
  0 & 1 &  1 & 1.9926 * & ------ & ------ &  0.40 & 0.80 \\
&&&&&&& \\
  0 & 2 &  0 & 2.2835 * & 2.2990 & 2.1937 &  0.20 & 0.30 \\
  0 & 2 &  1 & 2.6138   & ------ & ------ &  0.40 & 0.80 \\
  0 & 2 &  2 & 2.3222 * & ------ & ------ &  0.40 & 0.80 \\
&&&&&&& \\
  0 & 3 &  0 & 2.6574 * & ------ & ------ &  0.10 & 0.30 \\
  0 & 3 &  1 & 3.1665   & ------ & ------ &  0.40 & 0.80 \\
  0 & 3 &  2 & 2.8157   & ------ & ------ &  0.40 & 0.80 \\
&&&&&&& \\
  0 & 4 &  0 & 3.0675 * & ------ & ------ &  0.15 & 0.20 \\
  0 & 4 &  1 & 3.3175   & ------ & ------ &  0.40 & 0.80 \\
  0 & 4 &  2 & 3.3127   & ------ & ------ &  0.40 & 0.80 \\
\tableline
\cline{1-8}
  1 & 0 &  0 & 2.2567 * & 2.1483 & 2.1483 &  0.20 & 0.40 \\
&&&&&&& \\
  1 & 1 &  0 & 2.6477 * & 2.4728 & 2.4446 &  0.10 & 0.15 \\
  1 & 1 &  1 & 2.5909   & ------ & ------ &  0.40 & 0.80 \\
&&&&&&& \\
  1 & 2 &  1 & 3.0787   & ------ & ------ &  0.40 & 0.80 \\
  1 & 2 &  2 & 2.8879   & ------ & ------ &  0.40 & 0.80 \\
\tableline
\cline{1-8}
  2 & 0 &  0 & 2.9298 * & 2.6707 & 2.6707 &  0.10 & 0.20 \\
\end{tabular}
\label{tab2}
\end{table}
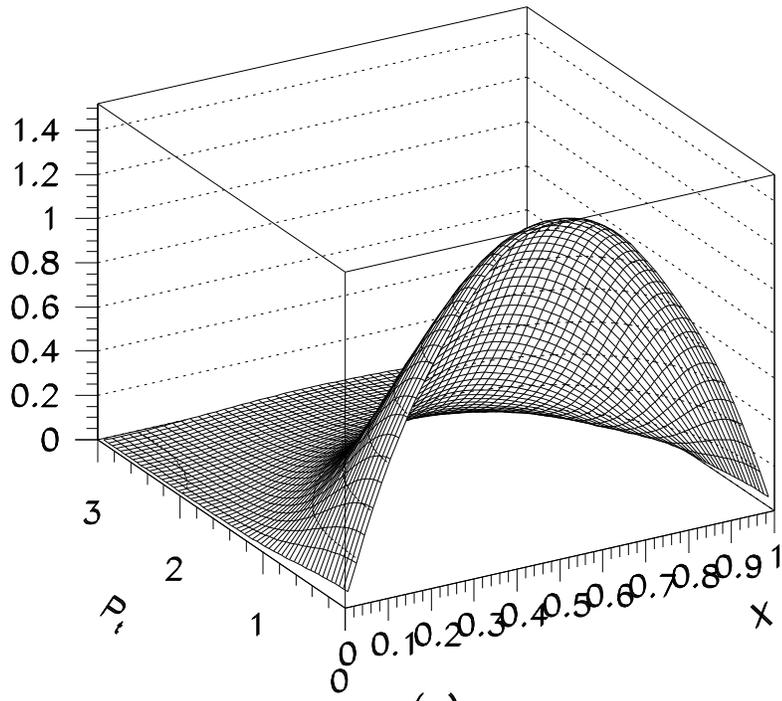
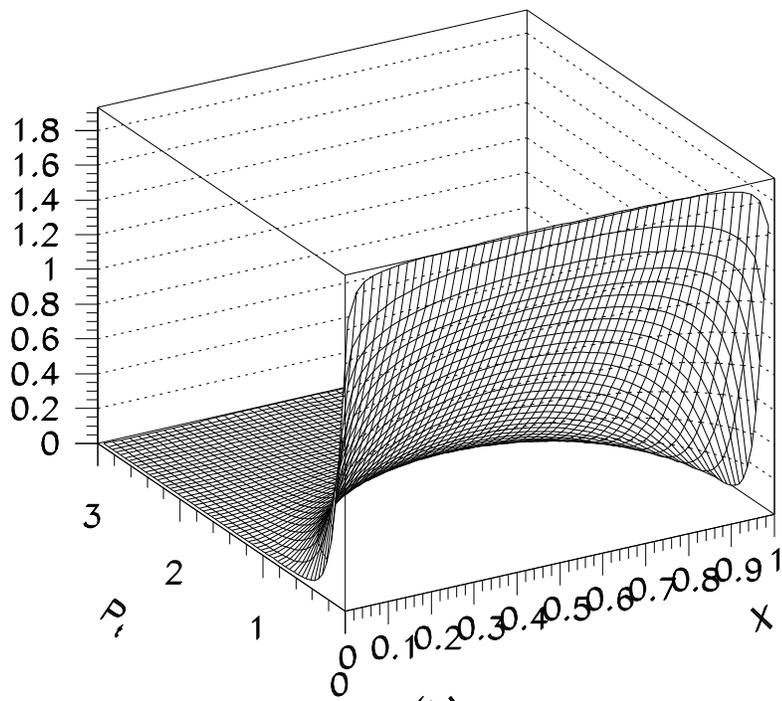
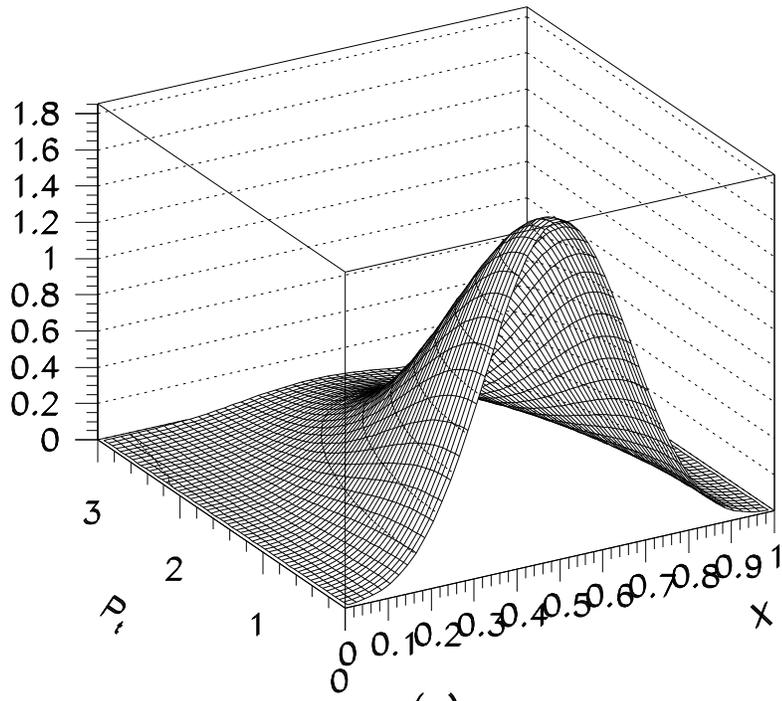
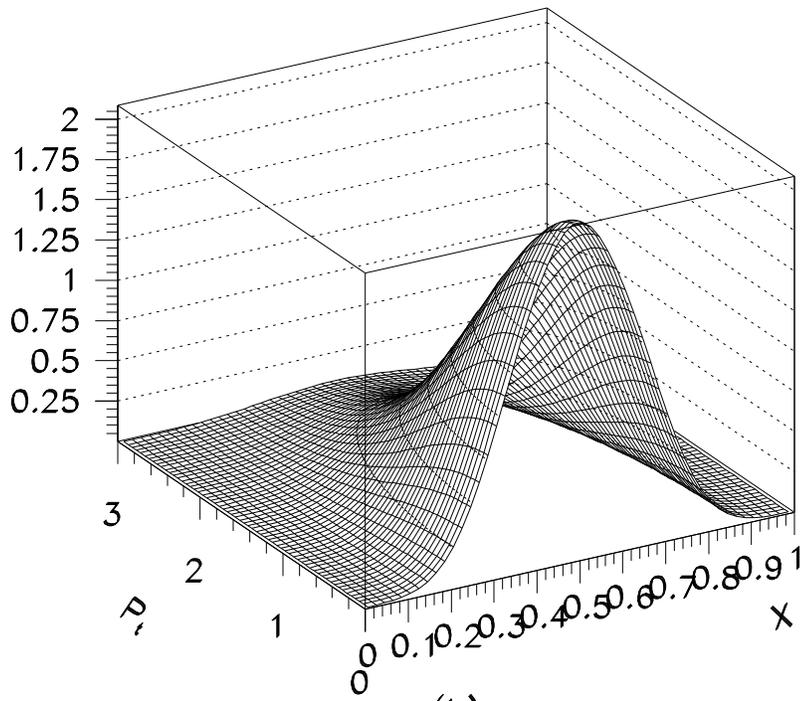
\end{document}